# E1 and E2 contributions to the L$_3$ resonance line shape in antiferromagnetic holmium

**L Bouchenoire[1,2], A Mirone[3], S D Brown[1,2], P Strange[4], T Wood[3], P Thompson[1,2], D Fort[5] and J Fernández-Rodríguez[3]**

[1] XMaS, European Synchrotron Radiation Facility, 6 rue Jules Horowitz, BP220, 38043 Grenoble, France

[2] Department of Physics, University of Liverpool, Oliver Lodge Laboratory, Oxford Street, Liverpool, L69 7ZE, UK

[3] European Synchrotron Radiation Facility, 6 rue Jules Horowitz, BP220, 38 043 Grenoble, France

[4] School of Physical Sciences, University of Kent, Canterbury, Kent, CT2 7NH, UK

[5] School of Engineering, Metallurgy and Materials, University of Birmingham, Edgbaston, Birmingham, B15 2TT, UK

E-mail: boucheno@esrf.fr

**Abstract.** A detailed study of the angular, energy and polarization dependences of the electric dipolar (E1: 2$p$→5$d$) and quadrupolar (E2: 2$p$→4$f$) contributions to the x-ray scattering cross-section is presented for holmium in its basal plane spiral antiferromagnetic phase at the L$_3$ edge. The corresponding E1 and E2 scattering factors have been extracted from fits to the experimental energy line shapes taking into account for the first time a split dipole resonance. Using the imaginary part of the resonant scattering factors to retrieve the XMCD spectrum, we find qualitative agreement with the dichroic spectrum measured in transmission through a holmium foil.

PACS numbers: 75.50.Ee, 61.10.Dp, 78.70.Ck

**Contents**

**1. Introduction**

The magnetism of rare-earth (RE) materials is of huge technological importance. Among other key applications, REs can be found in high-performance permanent magnets [1-3], as dopants in optical data storage devices [4,5] and in high speed magnetoelectronic devices [6,7]. An improved understanding of the electronic and magnetic properties of RE metals, including their interactions with other atoms in compounds and alloys, is crucial for developing future technologies. To this end, synchrotron x-rays are proving to be an important probe of RE magnetism [8,9]. Such experiments are carried out close to an atomic absorption edge by studying either absorption – namely, x-ray magnetic dichroism, either with circularly (i.e. XMCD) or linearly (i.e. XMLD) polarized photons – or scattering, where scattering can be performed either in the elastic or inelastic regime, namely by x-ray resonant magnetic scattering (XRMS) or resonant inelastic x-ray scattering (RIXS) respectively.

Controversy, however, surrounds the identification of E1 (dipolar) and E2 (quadrupolar) features in energy line shapes measured across the RE $L_{2,3}$ absorption edges in XRMS [10-12], XMCD [13-16] and RIXS [17,18]. The observation of a pre-edge peak (where the position of the "absorption edge" itself is normally defined as the first inflection point of an absorption curve) has generally been interpreted as arising, purely, from E2 processes, whereas the higher energy peak has been attributed to E1 excitations. However, deviations from this understanding have appeared in the literature: in an XMCD study at the Yb $L_3$ absorption edge, where a dichroic signal observed *at the edge* was interpreted to be E2 in origin, while a lower energy feature was assigned to be E1 [19]; in an XRMS study at the Tb $L_{2,3}$ edges where (again) the low energy peak was interpreted to be E1 and the high energy peak to be E2 [20]. More importantly, a recent combined experiment and theory investigation of charge-magnetic x-ray resonant interference scattering (XRIS) – a technique which is a "subset" of *elastic* resonant scattering (i.e. XRMS) – carried out by Brown *et al* [21,22], studying ferro- and ferri-magnetic phases of the hexagonal close-packed, heavy RE metals (moments along the c-axis), has challenged the general interpretation concerning the identification of E1 and E1 features. Their study relied upon strongly reducing the E2 cross-section by scattering at a normal Bragg position around 90°, with the incident polarization in the scattering plane (this minimized the E2 contribution via the polarization dependence of its scattering amplitude [23]). At the $L_3$ edge, for Ho (scattering angle, $2\theta = 96.1°$), Er ($2\theta = 93.3°$) and Tm ($2\theta = 89.5°$) the expected E2 contribution (determined from the scattering amplitude) was reduced by a factor of 0.10, 0.06 and 0.01, respectively. In spite of this reduction, a strong peak lying at the pre-edge position was observed for each RE. Thus, the authors concluded that the low and high energy resonant peaks arose from E1 transitions, implying a splitting of the *d*-band polarization caused by *f-d* spin hybridization just above the Fermi energy. The results were also consistent with first principles band structure calculations.

Having established the split E1 resonances across the heavy RE, it is natural to attempt to quantify the relative E1 and E2 contributions. In the present paper we revise the interpretation of E1 and E2 features occurring in the $L_3$ line shape of the first antiferromagnetic (AFM) material to be studied by XRMS [10-12], viz holmium metal in its basal plane spiral phase. We demonstrate that by taking into account the low energy E1 resonance and by exploiting the angular (i.e. measuring at different AFM satellite positions), energy and polarization dependences of the XRMS cross-section, the E1 and E2 contributions can be disentangled and their relative amplitude determined. The amplitude of the E1 and E2 scattering factors are presented along with the energy dependences of their respective real and imaginary parts. In contrast with the work reported by Brown *et al.* [21], only pure magnetic scattering is presented in this article. Also, in the XRIS study of Ho [21], the energy dependence of the E1 contribution was studied at just one single reflection, i.e. the Bragg position with $2\theta$ closest to 90°.

XMCD sum-rules, which are different for E1 and E2, cannot be reliably employed with the assumption that the low energy feature is purely E2 as has generally been the case to date. An accurate quantification of the E1 contribution to this spectral feature is therefore essential for reliable applications of the XMCD sum-rules. In that respect, we show in this article that it is possible to recover the XMCD spectrum by using the imaginary part of the resonant scattering factors and thus to extract the energy dependence of both E1 and E2 contributions that compose it. Knowing the exact position of the pure electric E1 and E2 transitions can also aid understanding of new phenomena that originate from E1-E2 scattering in non-centrosymmetric systems [24] (to be addressed in the future).

The paper is organized as follows. The experimental details are summarized in section 2. The equations describing the resonant and non-resonant contributions to the total scattering cross section are outlined in section 3 for the case of Ho. The experimental results are presented in section 4. The procedure adopted to fit the experimental line shapes and the results are discussed in section 5 and 6, respectively. A complete description of the resonant E1 and E2 scattering factors (real and imaginary parts, relative amplitudes and relative energy positions) is also presented in section 6. Finally, we demonstrate in section 6 how the XMCD spectrum can be recovered using the imaginary parts of the E1 and E2 scattering factors.

## 2. Experimental details

The measurements were carried out on the XMaS beamline [25] at the European Synchrotron Radiation Facility located on the soft end of the bending magnet BM28. A double bounce Si(111) monochromator with an energy resolution of 1.1 eV was tuned to the Ho $L_3$ edge. A toroidal mirror focused the monochromatic beam at the sample position to less than 1 mm$^2$. The vertical and horizontal slits were set such that the horizontal divergence was 3.5 times larger than the vertical divergence from the bending magnet source. The Ho single crystal grown at the University of Birmingham was a 5 mm x 15 mm, (001) surface cut. The rocking width was 0.02° on the (006). The 6-circle Huber diffractometer allowed rapid switching from horizontal scattering (so called π incident polarization) to vertical scattering (σ incident polarization) geometries.

Below the ordering temperature $T_N \approx 132K$, Ho enters a spiral antiferromagnetic phase in which the moments are confined to the *ab* plane in ferromagnetic sheets but rotate from plane to plane along the *c*-axis. The subsequent antiferromagnetic modulation wave vector, τ, decreases with decreasing temperature from 0.29c* at 132K down to 0.18c* at 13K before locking into a value of ~1/6c*, where a conical spiral structure exists with a ferromagnetic component along the c-axis.

The experiment was carried out on the (0 0 2n±τ) satellites with n =1,2 and 3. The sample was cooled to 40K with a closed cycle refrigerator leading to a value of τ =0.199(8)c*. The polarization analysis was achieved using the graphite (006) reflection of 0.3° rocking width.
All the resonant line shapes were obtained by measuring energy scans whilst monitoring the scattered polarization channel, such that one could define σ→σ, σ→π, π→σ and π→π scattering channels. The energy scans illustrated in figure 1 were measured at the (006+τ) satellite position in σ→σ (squares) and σ→π (dots) keeping the same scattering vector (*k*) at each energy. By comparison, it took twenty times longer to measure the same spectra by rocking the polarization analyzer (PA) crystal at every energy point (empty squares and circles). Given that the later method did not significantly improved the quality of the data and was more time consuming, energy scans were performed at fixed-*k* position for the rest of the measurements.

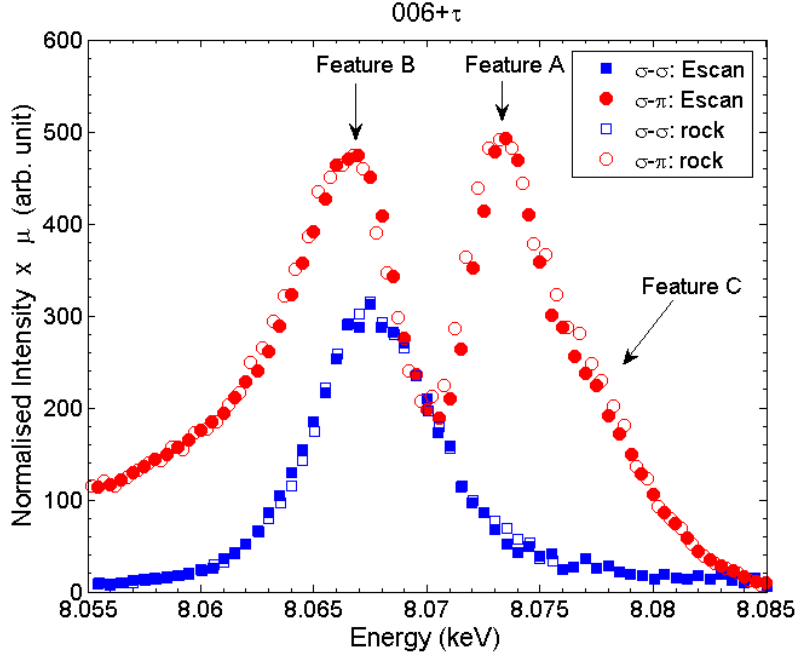

Figure 1. (Colour online) Comparison of two methods to measure energy line shapes at the Ho (006+τ) in σ→σ (squares) and σ→π (circles): standard energy scans keeping *k* constant (filled squares and dots), scans rocking the PA crystal at each energy (empty squares and circles).

## 3. Total x-ray scattering cross-section (TMCS)

The (0 0 2n±τ) reflections were analyzed in the theoretical framework of a spherical magnetic atom [23,24]. Only the final equations are presented in this section. A detailed review of the total scattering cross-section for magnetic moments lying in the *ab* plane can be found in Appendix A.
The observed intensity is proportional to the modulus squared of the total scattering amplitude. The total scattering amplitude of the ±τ satellites is the summation of the resonant, $F_{res}^{(mag)}$, and non-resonant contributions, $F_{non-res}^{(mag)}$, defined below in a matrix form in terms of the photon polarization:

$$F_{res}^{(mag)} = \begin{pmatrix} F_{\sigma\sigma} & F_{\pi\sigma} \\ F_{\sigma\pi} & F_{\pi\pi} \end{pmatrix} = iF_{E1}^{(1)} \begin{pmatrix} 0 & -c(\theta)/2 \\ c(\theta)/2 & i\,s(2\theta)/2 \end{pmatrix} \quad (1)$$

$$+ iF_{E2}^{(1)} \begin{pmatrix} i\,s(2\theta)/8 & -c(\theta)c(2\theta)/8 + 5s(\theta)s(2\theta)/16 \\ c(\theta)c(2\theta)/8 - 5s(\theta)s(2\theta)/16 & i\,7s(4\theta)/16 \end{pmatrix}$$

$$+ iF_{E2}^{(3)} \begin{pmatrix} i\,s(2\theta)/4 & -s(\theta)s(2\theta)/8 - c(\theta)c(2\theta)/4 \\ s(\theta)s(2\theta)/8 + c(\theta)c(2\theta)/4 & i\,s(4\theta)/8 \end{pmatrix}$$

$$F_{non-res}^{(mag)} = i\,\frac{\hbar\omega}{mc^2}\,\frac{s(2\theta)}{2} \begin{pmatrix} -i\,S_{eff} & s(\theta)(L_{eff} + S_{eff}) \\ -s(\theta)(L_{eff} + S_{eff}) & -i\,2L_{eff}\,s^2(\theta) - i\,S_{eff} \end{pmatrix} \quad (2)$$

where $\hbar\omega$ is the monochromator energy, $\theta$ is the sample Bragg angle, while $c$=$cos$ and $s$=$sin$ in shorthand notation. The values of the orbital and spin magnetization densities, $L_{eff}$ and $S_{eff}$, depend on the orbital and spin form factors, the Landé factor ($g$) and on the total angular momentum, $J$. The form factors were calculated (table 1) using values taken from [27] where fully relativistic Dirac-Fock wave functions were employed.

**Table 1.** Values of $2\theta$, $L_{eff}$ and $S_{eff}$ calculated for each Ho magnetic satellite at 8.072 keV.

| Satellite reflections | $2\theta$ (in °) | $L_{eff}$ | $S_{eff}$ |
|---|---|---|---|
| (006+τ) | 115.9 | 2.225 | 0.437 |
| (006-τ) | 104.9 | 2.381 | 0.488 |
| (004+τ) | 70.1 | 3.215 | 0.816 |
| (004-τ) | 62.6 | 3.451 | 0.920 |
| (002+τ) | 35.0 | 4.262 | 1.309 |
| (002-τ) | 28.5 | 4.406 | 1.389 |

The scattering factors $F_{E1}^{(1)}$, $F_{E2}^{(1)}$ and $F_{E2}^{(3)}$, which are complex functions, have the same definition as in [23]. For the data analysis (see section 5), they will be expressed as a sum of single oscillator functions, thus ensuring the energy dependence of their real and imaginary parts is consistent with their Kramers-Kronig Transformation.

As the measured intensity is proportional to the modulus squared of the total scattering amplitude, one can see that $\pi\rightarrow\sigma$ is equal to $\sigma\rightarrow\pi$. As a consequence, the same symbol (dots) and color (red) will be used for both channels in figures 2 and 3. Moreover, equation (1) shows that the $\sigma\rightarrow\sigma$ channel only contains contributions from the E2 (no E1) whereas both E1 and E2 occur in the other channels. This will be important for the data modeling in section 5.

## 4. Experimental results

The resonant line shapes presented in figures 2 and 3 were corrected for the sample absorption [28] and the Lorentz factor as follows:

$$I_{correct} = I_{meas}\, 2\mu \sin 2\theta \qquad (3)$$

The absorption correction was determined from the transmission measured through a 5μm thick polycrystalline Ho foil which was rescaled to tabulated values below and above the absorption located at $\hbar\omega$=8.072 keV.

The data were not corrected for the leak through between the un-rotated ($\sigma\rightarrow\sigma$ and $\pi\rightarrow\pi$) channels and the rotated ($\sigma\rightarrow\pi$ and $\pi\rightarrow\sigma$) channels as this was found to be negligible (<3%) during our measurements.

The $\pi\rightarrow\pi$ (triangles in figure 2) and $\pi\rightarrow\sigma$ (dots in figure 2) data were also multiplied by a factor of 3.5 corresponding to the difference between the horizontal and vertical divergence of the x-ray beam. The subsequent values obtained in $\pi\rightarrow\sigma$ are close to those of the $\sigma\rightarrow\pi$ channel as expected from the expression for the total cross-section.

For a given scattering geometry, the line shape always shows two peaks for each satellite except for $\sigma\rightarrow\sigma$ (squares in figure 3). Interestingly, an XRMS study on terbium at the $L_{2,3}$ edges *did not* report any signal in the $\sigma\rightarrow\sigma$ channel, and thus an absence of E2 scattering was concluded [29]. It should be mentioned, however, that the "large background" reported in that work could have limited the observation of any E2 contribution.

For the (006+τ), the main high energy peak (Feature A, figure 2) is found 6±0.5 eV above the low energy peak (Feature B, figure 2). For comparison, Gibbs *et al* [12] measured the (0 0 2n+τ) with $n$ =1,2 and 3 in $\sigma\rightarrow\pi$ but the reported energy resolution was insufficient (5-10 eV at the $L_3$ edge) to resolve the two peaks. In our data, the separation between the two peaks decreases (dashed lines in figure 2) for lower photon momentum transfer, and is about 3.5±1.0 eV for the (002±τ).

All the data except those measured in the $\sigma\rightarrow\sigma$ channel show a high energy shoulder (Feature C, figure 2) which appears ~3eV above Feature A.

We note that the relative sizes of the (002+τ) features in π→π and π→σ are comparable to those reported by Grübel *et al* [30].

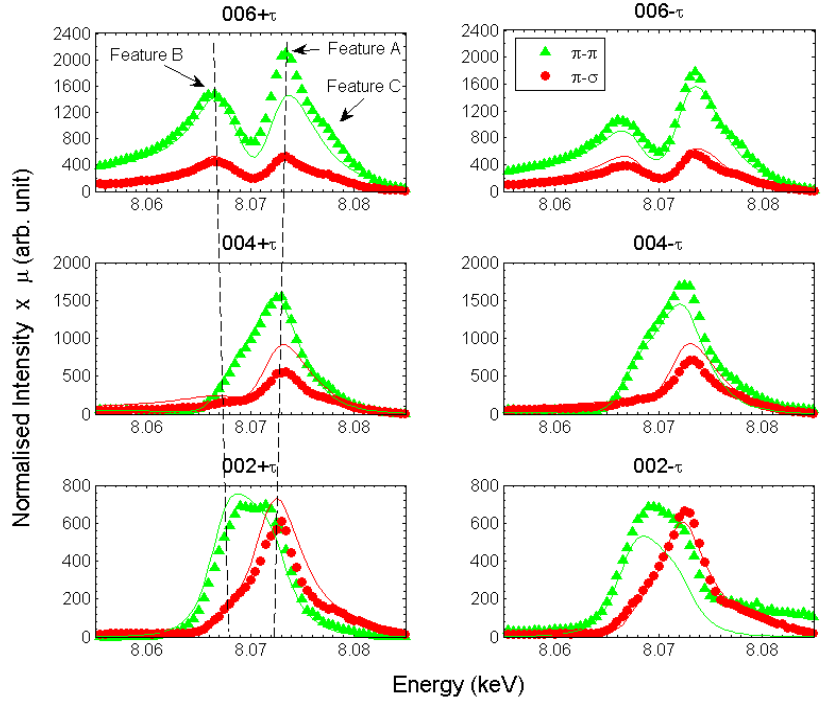

**Figure 2.** (Colour online) Measured resonant line shapes across the Ho $L_3$ edge in π→π (triangles), π→σ (dots) and comparison to the fits (solid lines) described in the text. The dashed lines are only guides to the eye. They show the energy contraction between Features A and B.

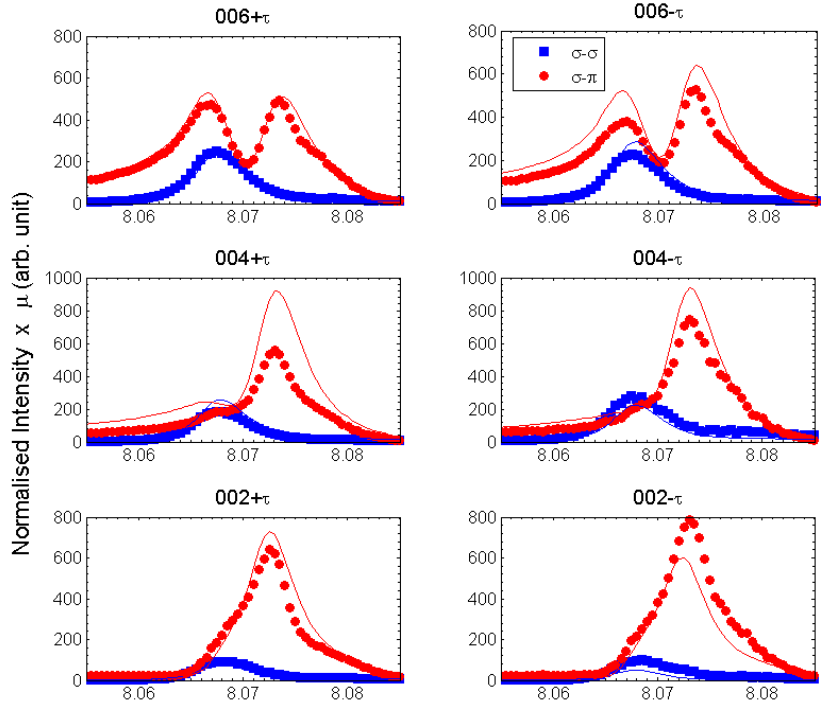

**Figure 3.** (Colour online) Measured resonant line shapes across the Ho $L_3$ edge in σ→σ (squares), σ→π (dots) and comparison to the fits (solid lines) described in the text.

## 5. Fitting program outputs

The fitting procedure is described in detail in Appendix B. The refinement was achieved fitting simultaneously the twenty four experimental line shapes illustrated in figures 2 and 3. The resulting fits (solid lines in figures 2 and 3) were obtained by modeling [31] the imaginary and real parts of the scattering factors ($F_{E1}^{(1)}$, $F_{E2}^{(1)}$ and $F_{E2}^{(3)}$) described as the sum of single oscillator functions. The E1 term, $F_{E1}^{(1)}$, was modeled with three oscillator functions while two oscillator functions were used for each E2 term ($F_{E2}^{(1)}$ and $F_{E2}^{(3)}$). However, we stress that the choice of the number of oscillator functions is purely phenomenological, representing the minimum number required to fit the experimental data. Given that all the data except for σ→σ exhibit three features (A, B and C) and knowing from [21] that the low energy feature contains an E1 contribution, three was the minimum number of oscillator functions to fit $F_{E1}^{(1)}$. According to equation (1), σ→σ is the only channel that contains exclusively $F_{E2}^{(1)}$ and $F_{E2}^{(3)}$. This channel displays a slightly asymmetric peak, which clearly cannot be described by a single oscillator. Thus, it was described by a minimum of two oscillators for each E2 term. Note that the E2 multiplet structure, which has been observed for example by Tanaka *et al* [32], cannot be resolved with the energy resolution obtained in the measurements reported here.

A least-squares method was utilized to minimize the Normalized Squared Error (*NSE*) defined in Appendix B. The outputs of the code are summarized in tables 2 to 5.

**Table 2.** Outputs of the fitting program describing $F_{E1}^{(1)}$.

| $h_{11\ a}$ (arb. u.) | $h_{11\ b}$ (arb. u.) | $h_{11\ c}$ (arb. u.) | $E_{11\ a}$ (keV) | $E_{11\ b}$ (keV) | $E_{11\ c}$ (keV) | $\gamma_{11\ a}$ (eV) | $\gamma_{11\ b}$ (eV) | $\gamma_{11\ c}$ (eV) |
|---|---|---|---|---|---|---|---|---|
| 1.6 | 3.8 | -8.3 | 8.0713 | 8.0715 | 8.0735 | 4.2 | 2.6 | 4.0 |

**Table 3.** Outputs of the fitting program describing $F_{E2}^{(1)}$.

| $h_{21\ a}$ (arb. u.) | $h_{21\ b}$ (arb. u.) | $E_{21\ a}$ (keV) | $E_{21\ b}$ (keV) | $\gamma_{21\ a}$ (eV) | $\gamma_{21\ b}$ (eV) |
|---|---|---|---|---|---|
| -8.8 | 12.4 | 8.0665 | 8.0670 | 3.4 | 3.4 |

**Table 4.** Outputs of the fitting program describing $F_{E2}^{(3)}$.

| $h_{23\ a}$ (arb. u.) | $h_{23\ b}$ (arb. u.) | $E_{23\ a}$ (keV) | $E_{23\ b}$ (keV) | $\gamma_{23\ a}$ (eV) | $\gamma_{23\ b}$ (eV) |
|---|---|---|---|---|---|
| -4.6 | 6.6 | 8.0665 | 8.0670 | 3.4 | 3.4 |

**Table 5.** Remaining outputs of the fitting program.

| *G1* (arb. u.) | *DL* (μm) | 2σ (eV) | *G2* (arb. u.) |
|---|---|---|---|
| 316 | 0.36[a] | 1.1 | 17 |

[a] "*DL*" is expressed in "μm" in table 5 but is used in "cm" in the computer code (i.e. "*DL*=3.6e-5")

The width of each oscillator (=2 x $\gamma_{LN\_t}$ where $\gamma_{LN\_t}$ is the half-width of the oscillator, see Appendix B) is larger than the natural width semi-empirical value of 4.26±0.43 eV estimated for Ho at the $L_3$ level [33]. This broadening of the width is partially due to the convoluted effect of the core hole life time and the width of the intermediate (excited) states.

## 6. Discussion

In this section, we firstly describe (section 6.1) how the fits successfully reproduce the majority of the features present in the experimental line shapes. We make suggestions about the physical meaning of the high energy shoulder present in the data but not reproduced by the fits. We then discuss the quality of the fits in terms of the *NSE*. In section 6.2, we describe the shape, the amplitude of the resonant scattering factors and their respective real and imaginary parts. Finally, we show in section 6.3 how the dichroism spectrum can be recovered using the imaginary part of the resonant scattering factors. The energy position of the E1 and E2 contributions to the XMCD spectrum is discussed and compared with published results.

*6.1. Quality of the fits*

The fits are in good agreement with the data. The asymmetric tail especially present in the (006±τ) could only be reproduced by including the non-resonant part of the cross-section, $F_{non-res}^{(mag)}$, in the computer code [31]. This asymmetry results from the interference between the resonant and non-resonant contributions to the cross-section as observed in [12]. A similar but smaller asymmetry could also be reproduced in the high energy side of the σ→σ spectra.
The fits also reproduced the energy contraction between Feature A and B shown as the dashed lines in figure 2. We concluded that this contraction was only due to the angular terms in equation (1). To establish this, we plotted the modeled curves by feeding all the parameters back into (A.1) except the parameters "*G2*" and the dead layer, "*DL*", in order to evaluate the effect of i) the non-resonant contribution (by setting "*G2*=0" and "*DL*=3.6e-5"), ii) the dead-layer (by setting "*DL*=0" and "*G2*=17") and iii) both the non-resonant contribution and the dead-layer together (by setting "*G2*=0" and "*DL*=0"). The energy contraction between Feature A and B was still observed for the three cases proving that this effect could only result from the $\theta$ dependence of the scattering factors in equation (1).
Although the fits successfully reproduced the peaks (Features A and B) and the asymmetric tails, they did not show any evidence of the presence of the shoulder depicted as Feature C, which is located ~2eV above the white line. The data however suggest that this feature is dipolar in nature as it is present in all the channels except in σ→σ. Recent band structure calculations based on the SIC-LDA [34,21] suggest strong 5*d*-6*s* hybridization in the high energy tails of the spectra, which may account for Feature C.
The results of our fits (using 3 and 2 oscillators, respectively for $F_{E1}^{(1)}$ and $F_{E2}^{(1,3)}$) gives a *NSE* of 0.06. Adding more oscillator functions does not significantly improve the *NSE*. For example, fitting four oscillators for $F_{E1}^{(1)}$ instead of three improves (reduces) *NSE* by only 0.4% ; similarly, fitting $F_{E2}^{(1)}$ and $F_{E2}^{(3)}$ with three oscillator functions instead of two improves *NSE* by only 0.2%. On the other hand, if the low energy E1 feature is completely ignored, the use of only one oscillator function to fit the E1 contribution results in a 30% degradation (increase) in *NSE*. Given that twenty four energy scans (i.e. four scattering channels for six reflections) were fitted simultaneously, satisfactory agreement between the fits and the experimental data is obtained. One way of improving the fits could be to parameterize one scaling factor for each energy line shape instead of using the global factor *G1* in (B.1). These scaling factors would take into account any experimental artifacts (inhomogeneous dead-layer thickness, slightly different illuminated areas, …). This would however increase the number of fitting parameters (25 in our present code).

*6.2. Resonant scattering factors*

The real and imaginary parts of each scattering factor are plotted in figures 4(a) and 4(b) as a function of energy. These scattering factors are independent of the scattering geometry. The imaginary and real parts of the two E2 scattering terms (figure 4(b)) are opposite in sign to those of $F_{E1}^{(1)}$ (figure 4(a)). This difference in sign between E1 and E2 has been attributed to the spin dependence of the dipole matrix elements. Various explanations to this sign difference between E1 and E2 have been given over the years: contraction/expansion or "breathing″ of the 5*d* orbitals [35,36], the spin-orbit coupling [37] and crystalline electric field effects [38]. A more recent paper [21] suggested that this difference in sign results from anti-parallel spin hybridization of the unoccupied 4*f*-5*d* states.
The energy dependence of the modulus of each scattering factor is presented in figure 5. The two E2 terms maximize at 8.0675 keV, i.e. 4.5 eV below the absorption edge (8.072 keV), which also corresponds to energy of Feature B in figures 2 and 3. The presence of E2 resonances below the absorption edge can be explained as follows: when the core hole is created, the highly localized unoccupied 4*f*-states which lie initially above the Fermi Energy, $E_F$, become more tightly bound thereby lowering the 4*f* energy and creating an E2 excitation below the absorption edge. The ratio $F_{E2}^{(3)}/F_{E2}^{(1)}$ is found to be 0.56 and close the free-ion value [39].

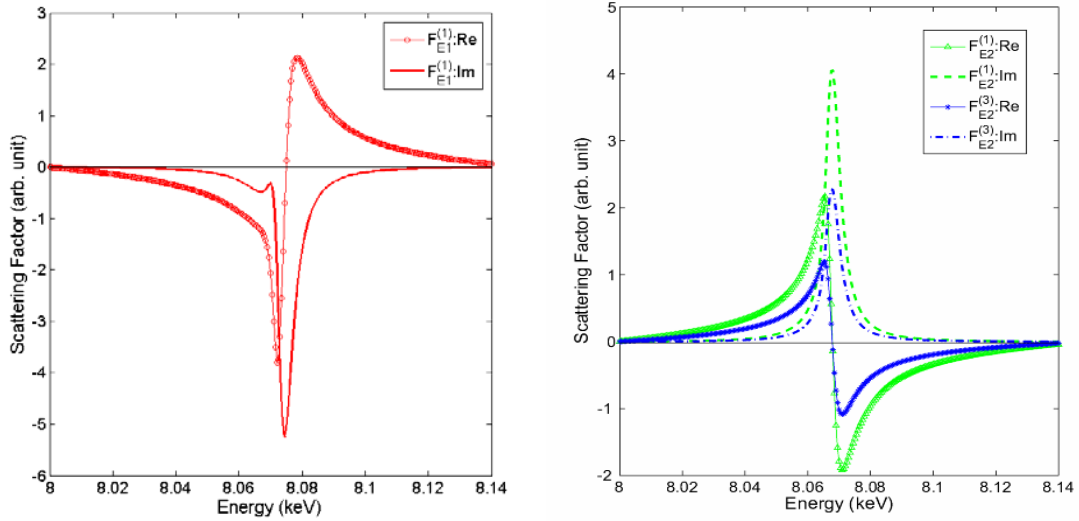

**Figure 4.** (Colour online) Real part of (a) $F_{E1}^{(1)}$ (circles), (b) $F_{E2}^{(1)}$ (triangles) and $F_{E2}^{(3)}$ (stars) plotted as a function of energy across the absorption edge and comparison with their respective imaginary part: (a) $F_{E1}^{(1)}$ (line), (b) $F_{E2}^{(1)}$ (dashed line) and $F_{E2}^{(3)}$ (dash-dotted line).

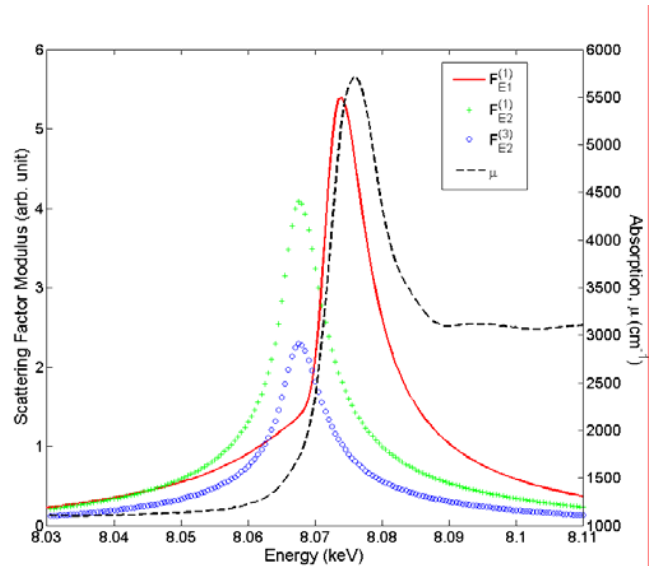

**Figure 5.** (Colour online) Scattering factor moduli: $F_{E1}^{(1)}$ (solid line), $F_{E2}^{(1)}$ (crosses) and $F_{E2}^{(3)}$ (circles). The dashed line is the measured absorption curve.

The most striking thing about figure 5 is that $F_{E1}^{(1)}$ (red line) exhibits a two-feature structure similar to the one observed in the ferromagnetic phase [21]. The main E1 peak maximizes ~6eV above the E2 peak and corresponds to the energy of Feature A in figures 2 and 3. It is also 1.5 eV above the absorption edge. The low energy E1 feature, better described as a shoulder, resonates at approximately the energy of the E2 contributions (8.068 keV±1.5eV). This peak results from hybridization between the 5*d* and 4*f* empty states [21]. As the 4*f* unoccupied levels are pulled down below $E_F$ due to the core hole, one can also expect a reduction of the energy of the 5*d* bands due to the *f-d* hybridization. First principles band theory [34] has also predicted a split *d*-band [21], with the lower energy E1 peak coinciding with the E2 contribution although this prediction was made for a c-axis moment. However, the code employed to calculate the predicted split *d*-band is unable to model the basal plane moments present in the experiment and so numerical agreement with experiment cannot be expected.

The ratio between the two E1 features is ~4:1. The close proximity to the white line strongly affects the relative amplitude between these two E1 features. The low energy E1 feature cannot be neglected as it represents ~ ⅓ of $F_{E2}^{(1)}$ (crosses in figure 5) and ~ ⅔ of $F_{E2}^{(3)}$ (circles in figure 5). Feature B in the resonant spectra of figures 2 and 3 therefore results not only from E2 processes but also from E1 excitations.

*6.3. Extraction of the XMCD information using the XRMS data*

Following the optical theorem which states that the absorption coefficient is directly proportional to the imaginary part of the forward scattering amplitude [40], one can expect that the dichroic spectrum across the Ho $L_3$ edge can be retrieved from the imaginary part of both E1 and E2 scattering factors extracted from the measured XRMS spectra. Carra and co-workers [41] described the angular dependence of the E1 and E2 contributions to circular dichroism in terms of transitions probabilities, $w_{(LM)}$ (see equations (7) and (8) in [41]). These expressions are analogous to those expressed for scattering in terms of the transitions probabilities $F_{LM}$ factors [11,23]. In turn, the resonant scattering amplitudes, $F_{EL}^{(N)}$, (here being $F_{E1}^{(1)}$, $F_{E2}^{(1)}$ and $F_{E2}^{(3)}$) are combinations of the complex quantities $F_{LM}$ (see (A.8) and (B.2)). The total XMCD spectrum obtained after substitution of the $w_{(LM)}$ terms by the imaginary part of the $F_{EL}^{(N)}$ measured in our XRMS experiment is presented as the solid line in figures 6 and 7.

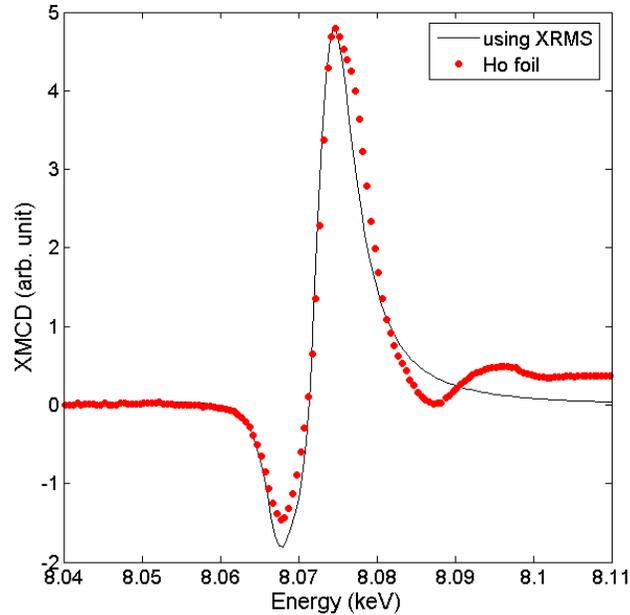

**Figure 6.** (Colour online) XMCD spectrum obtained using the imaginary part of the resonant scattering terms (solid line). The dots represent the dichroic spectrum measured in transmission through a 5μm thick Ho foil by reversing the photon helicity under a constant magnetic field and at 10K. The overall spectrum using the XRMS data was multiplied by -1 to agree with the sign of that measured in transmission with the foil.

In figure 6, the total XMCD spectrum is compared to the dichroic spectrum measured in transmission through the same polycrystalline foil (dots) as that used for the absorption correction of our XMRS experiment. This former spectrum was measured at 10K by reversing the photon helicity under a 1T magnetic field applied along the incident photon direction. The helicity was reversed from left to right at 11.5Hz using the XMaS phase-plate flipper [42] and the dichroic spectrum was obtained using lock-in detection [43]. The overall XMCD spectrum obtained using the XRMS imaginary parts was multiplied by -1 to agree with the sign of the spectrum measured with the foil (the sign of the oscillators in the XRMS measurements was arbitrary). As both spectra were rescaled, only qualitative comparison is possible. Given that the two spectra were measured for two different sample types (single crystal/foil) and at two different temperatures (40K/10K), there is good qualitative agreement between them. The XMCD spectrum extracted from the XRMS data does not however reproduce the XANES oscillations above the white line. The E1 (line with crosses) and E2 (dash-dotted line) contributions to the total dichroic spectrum are presented in figure 7. The positive peak is mainly E1 whereas the negative peak originates from both E2 and E1 contributions. Recent circular dichroism results performed at the $L_3$ edge of Gd, Tb and Dy [16] also supported the idea of an overlap between the E1 and E2 spectra a few eV below the edge. However, the shape, position and sign of the E2 contribution relative to the E1 part is so different between Tb and Dy that it is not possible to deduce [16] what the results should be for the case of Ho. In that respect, our XMCD results obtained with a completely different method gives a qualitative indication of the energy dependence of both E1 and E2 contributions with respect to each other.

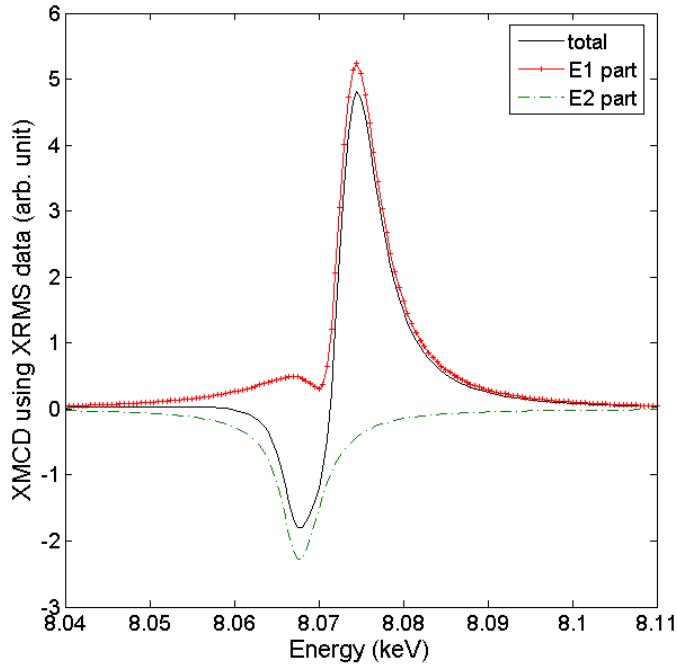

**Figure 7.** (Colour online) XMCD spectrum obtained using the imaginary part of the resonant scattering terms (solid line) along with the E1 (line with the crosses) and E2 parts (dash-dotted line) to the total dichroic spectrum.

## 7. Conclusion

In conclusion, the angular, energy and polarization dependences of the E1 and E2 contributions to the scattering cross-section have been measured in detail in the basal-plane spiral antiferromagnetic phase of Ho. The E1 and E2 scattering terms were extracted individually by fitting single oscillator functions to the data and taking into account for the first time a double E1 feature. The fits unambiguously confirm that the low energy peak observed in the resonant energy line shapes of Ho originates from mixed E1 and E2 excitations and that the E1 contribution is non negligible. The origin of the high energy shoulder (Feature C) still has to be clarified. The results demonstrate that a complete polarization analysis of the XRMS is a valuable technique which allows accurate identification of all the resonance processes and quantification of each scattering term that contributes. Using the imaginary part of the resonant scattering factors obtained with our phenomenological approach, we reconstructed the XMCD spectrum across the Ho $L_3$ edge. Our results are in good qualitative agreement with the dichroic spectrum obtained in transmission through a foil. These results, in conjunction with those in [21,22], also suggest that the XMCD spectra measured across the heavy RE metals and alloys, which assumed that the low energy peak is purely E2, may need to be readdressed.

The study of the E1 and E2 angular, energy and polarization dependences will be repeated at the $L_2$ edge in the near future. We will be able to compare their relative sizes at the $L_2$ edge and also between the $L_2$ and $L_3$ edges. Finally, we anticipate that the new spectroscopic interpretations obtained using RXMS and XRIS will lead to better application of sum rules to accurately quantify the spin and orbital moments for the $L_2$ and $L_3$ absorption edges in rare-earths materials.

**Acknowledgements**

We gratefully acknowledge the financial support of EPSRC. The experimental work was performed on the EPSRC-funded XMaS beamline at the ESRF, directed by M.J. Cooper and C.A. Lucas. One of us, J.F.R., is grateful to Gobierno del Principado de Asturias for the financial support of a Postdoctoral grant from Plan de Ciencia, Tecnologia e Innovacion (PCTI) de Asturias 2006-2009

**Appendix A. Derivation of the TMCS for magnetic moments lying in the a-b plane**

The holmium (0 0 $2n\pm\tau$) reflections were analyzed in the theoretical framework of a spherical magnetic atom [23,24] (the non-centrosymmetric part does not contribute in that case). In such a framework the resonant scattering amplitude is written as follows considering E1-E1 and E2-E2 elements only:

$$F_{\varepsilon k \to \varepsilon' k'} = \frac{3}{2k}\sum_{M=-1}^{M=1} F_{1,M} \varepsilon_M^{1*} \varepsilon_M^1 + \frac{5}{k}\sum_{M=-2}^{M=2} F_{2,M} (k\varepsilon)_M^* (k\varepsilon)_M \tag{A.1}$$

where $\underline{k}$ ($\underline{k}'$) and $\underline{\varepsilon}$ ($\underline{\varepsilon}'$) are the incident (scattered) wave-vector and incident (scattered) polarization vector, respectively. For E1 (first term), $\varepsilon_M^1$ represents $\underline{\varepsilon}$ in rank 1 spherical tensor form and for E2 (second term), $(k\varepsilon)_M$ denotes the rank 2 spherical tensor components of $\underline{k}$ x $\underline{\varepsilon}$. The angular momentum quantization axis, $\underline{\xi}$, is taken parallel to the local magnetization. In the spherical case, the dependence in $M$ of the $F_{1,M}$ and $F_{2,M}$ which are the $F_{LM}$ factors described in [23], is entirely magnetic.

Experimentally, the measured intensity is proportional to the modulus squared of the scattering amplitude. For magnetic moments lying in the *ab* plane, (A.1) is independent of the sign of the magnetic field (i.e. if the magnetization of every Ho atom is reversed, (A.1) remains the same) and can therefore be simplified. This implies that the $n\pm p\tau$ reflections with $p$ even depend on the time-reversal symmetric part of the $F_{LM}$ factors while the $p$ odd reflections are given exclusively by their time-reversal antisymmetric part. In our case for the $2n\pm\tau$ reflections, only the time reversal antisymmetric part of the scattering amplitude is relevant and can be expressed as follows:

$$\begin{aligned} F_{\varepsilon k \to \varepsilon' k'} = &+i\frac{3}{4k}(F_{1,1}-F_{1,-1})(\varepsilon'_{x_m}\varepsilon_{y_m} - \varepsilon'_{y_m}\varepsilon_{x_m}) \\ &+i\frac{5}{4k}(F_{2,1}-F_{2,-1})\left[(k'\varepsilon')_{x_m z_m}(k\varepsilon)_{y_m z_m} - (k'\varepsilon')_{y_m z_m}(k\varepsilon)_{x_m z_m}\right] \\ &+i\frac{5}{4k}(F_{2,2}-F_{2,-2})\left[(k'\varepsilon')_{x_m^2-y_m^2}(k\varepsilon)_{x_m y_m} - (k'\varepsilon')_{x_m y_m}(k\varepsilon)_{x_m^2-y_m^2}\right] \end{aligned} \tag{A.2}$$

In the above expression the E1 and E2 terms are expressed in a rotating Cartesian framework $x_m$, $y_m$, $z_m$ with the $z_m$ axis aligned with the magnetization. For a spiral structure having the modulation vector $\tau$ lying along the (002n) direction, the magnetic moment takes the form:

$$\hat{z}_m = c(\tau . r_m)\hat{x}_m + s(\tau . r_m)\hat{y}_m \tag{A.3}$$

where $r_m$ is the position of the $m^{th}$ atom, while $c=cos$ and $s=sin$ in shorthand notation.
The scattering amplitude for the four possible scattering channels becomes, in a matrix form:

$$\begin{pmatrix} F_{\sigma\sigma} & F_{\pi\sigma} \\ F_{\sigma\pi} & F_{\pi\pi} \end{pmatrix} = i\frac{3}{4k}(F_{1,1}-F_{1,-1})\begin{pmatrix} 0 & -c(\theta)c(\tau.r_m) \\ c(\theta)c(\tau.r_m) & s(2\theta)s(\tau.r_m) \end{pmatrix}$$

$$+i\frac{5}{4k}(F_{2,1}-F_{2,-1})\begin{pmatrix} s(2\theta)s(\tau.r_m)c(2\tau.r_m) & -s(\theta)s(\tau.r_m)s(2\theta)s(2\tau.r_m)/2 \\ s(\theta)s(\tau.r_m)s(2\theta)s(2\tau.r_m)/2 & -c(\theta)c(2\tau.r_m)c(2\theta)c(\tau.r_m) \\ +c(\theta)c(2\tau.r_m)c(2\theta)c(\tau.r_m) & -c(2\theta)c(\tau.r_m)s(2\theta)s(2\tau.r_m) \end{pmatrix} \quad (A.4)$$

$$+i\frac{5}{4k}(F_{2,2}-F_{2,-2})\begin{pmatrix} -s(2\theta)s(2\tau.r_m)c(\tau.r_m)/2 & -c(\theta)s(2\tau.r_m)c(2\theta)s(\tau.r_m)/2 \\ c(\theta)s(2\tau.r_m)c(2\theta)s(\tau.r_m)/2 & +s(\theta)c(\tau.r_m)s(2\theta)[1+s(\tau.r_m)^2]/2 \\ -s(\theta)c(\tau.r_m)s(2\theta)[1+s(\tau.r_m)^2]/2 & -s(2\theta)[1+s(\tau.r_m)^2]c(2\theta)s(\tau.r_m) \end{pmatrix}$$

$$F_{non-res}^{(mag)} = i\frac{\hbar\omega}{mc^2}s(2\theta)\begin{pmatrix} s(\tau.r_m)S_{eff} & s(\theta)c(\tau.r_m)(L_{eff}+S_{eff}) \\ -s(\theta)c(\tau.r_m)(L_{eff}+S_{eff}) & s(\tau.r_m)(2s^2(\theta)L_{eff}+S_{eff}) \end{pmatrix} \quad (A.5)$$

where $\theta$ is the sample Bragg angle. The non-resonant contribution to the scattering amplitude, $F_{non-res}^{(mag)}$, can also be described in a matrix form [26] as follows:

where $L_{eff} = L_{eff}(K) = J\left(\frac{2-g}{g}\right)f_L(K)$ (A.6)

and $S_{eff} = S_{eff}(K) = J\left(\frac{g-1}{g}\right)f_S(K)$ (A.7)

Here $\underline{K}$ is the scattering vector such that $\underline{K} = \underline{k} - \underline{k}'$. The total scattering amplitude of the $\pm\tau$ satellites is the summation of the resonant, $F_{res}^{(mag)}$, and non-resonant contributions, $F_{non-res}^{(mag)}$:

$$F_{res}^{(mag)} = iF_{E1}^{(1)}\begin{pmatrix} 0 & -c(\theta)/2 \\ c(\theta)/2 & is(2\theta)/2 \end{pmatrix}$$

$$+iF_{E2}^{(1)}\begin{pmatrix} is(2\theta)/8 & -c(\theta)c(2\theta)/8+5s(\theta)s(2\theta)/16 \\ c(\theta)c(2\theta)/8-5s(\theta)s(2\theta)/16 & i7s(4\theta)/16 \end{pmatrix} \quad (A.8)$$

$$+iF_{E2}^{(3)}\begin{pmatrix} is(2\theta)/4 & -s(\theta)s(2\theta)/8-c(\theta)c(2\theta)/4 \\ s(\theta)s(2\theta)/8+c(\theta)c(2\theta)/4 & is(4\theta)/8 \end{pmatrix}$$

$$F_{non-res}^{(mag)} = i\frac{\hbar\omega}{mc^2}\frac{s(2\theta)}{2}\begin{pmatrix} -iS_{eff} & s(\theta)(L_{eff}+S_{eff}) \\ -s(\theta)(L_{eff}+S_{eff}) & -i2L_{eff}s^2(\theta)-iS_{eff} \end{pmatrix} \quad (A.9)$$

where the elements of the matrices are defined in terms of the photon polarization. The values of the orbital and spin magnetization densities, $L_{eff}$ and $S_{eff}$, depend on the orbital and spin form factors, the Landé factor ($g$) and on the total angular momentum, $J$.

**Appendix B. Fitting procedure**

The fitting procedure can be summarized by the following expression:

$$I_{fit}(\hbar\omega) \propto |Fit_{tot}|^2 = G1\times\int\left(\left|F_{res}^{(mag)}(\hbar\omega')+G2\times F_{non-res}^{(mag)}(\hbar\omega')\right|^2 \times e^{-2\mu\times DL/\sin\theta}\right)$$

$$\times e^{-(\hbar(\omega-\omega'))^2/2\sigma^2} \times d\omega \quad (B.1)$$

where $F_{res}^{(mag)}$ and $F_{non-res}^{(mag)}$ are the resonant and non-resonant contributions to the total scattering amplitude expressed in (A.8) and (A.9), respectively.

Each resonant scattering factor ($F_{E1}^{(1)}$, $F_{E2}^{(1)}$ and $F_{E2}^{(3)}$) was modeled as the sum of single oscillator functions as follows:

$$F_{EL}^{(N)}(\hbar\omega) = \sum_{EL\_t} F_{EL\_t}^{(N)}(\hbar\omega) = \sum_{LN\_t} \frac{h_{LN\_t} \times \gamma_{LN\_t}}{d_{LN\_t} - i\gamma_{LN\_t}} \tag{B.2}$$

where $d_{LN\_t} = E_{LN\_t} - \hbar\omega$

In our notation $N$ is the order number and $L$ represents the change in orbital angular momentum. In

$$F_{E1\_a}^{(1)}(\hbar\omega) = \frac{h_{11\_a} \times \gamma_{11\_a}}{(E_{11\_a} - \hbar\omega) - i\gamma_{11\_a}} \tag{B.3}$$

the spiral phase of Ho, $N=1$ and $L=1$ for E1 while $N=1$ and 3, and $L=2$ for E2. The subscript $t$ ($t=a, b, etc...$) labels the oscillators: e.g. the first oscillator ($F_{E1\_a}^{(1)}$) describing the E1 scattering factor $F_{E1}^{(1)}$ is expressed as follows:

The E1 scattering factor is $F_{E1}^{(1)} = F_{E1\_a}^{(1)} + F_{E1\_b}^{(1)} + $ etc...

The expression for the real and imaginary parts of each oscillator function is given by:

$$\mathrm{Re}\left(F_{EL}^{(N)}(\hbar\omega)\right) = \frac{h_{LN\_t} \times d_{LN\_t} \times \gamma_{LN\_t}}{d_{LN\_t}^2 + \gamma_{LN\_t}^2} \tag{B.4}$$

$$\mathrm{Im}\left(F_{EL}^{(N)}(\hbar\omega)\right) = \frac{h_{LN\_t} \times \gamma_{LN\_t}^2}{d_{LN\_t}^2 + \gamma_{LN\_t}^2} \tag{B.5}$$

Note that (B.5) describing the imaginary part of the oscillator has the form of a Lorentzian function.

The position ($E$), the height ($h$) and the half-width ($\gamma$) of each oscillator as well as the dead-layer thickness ($DL$) and the scaling factors ($G1$ and $G2$) were parameterized [26]. The global scale factor, $G1$, incorporates the structure factor, the Debye-Waller factor and the incident intensity. The convolution was carried out using a Gaussian function of 1.1eV width ($2\sigma$) to represent the energy resolution of the instrument at the Ho L$_3$ edge.

The twenty four fits (lines in figures 2 and 3), which represent the sum of both E1 and E2 contributions, were simultaneous achieved by modeling the imaginary and real parts of each resonant scattering factor.

The best refinement was obtained using least squares to minimize the Normalized Squared Error (*NSE*) defined in our code [31] as follows:

$$NSE = \frac{\sum_{i=1}^{i=24} \int (I_{fit}(\hbar\omega) - I_{exp}(\hbar\omega))^2 d\omega}{\sum_{i=1}^{i=24} \int (I_{exp}(\hbar\omega))^2 d\omega} \tag{B.6}$$

where $I_{fit}(\hbar\omega)$ is the modeled scattered intensity defined in (B.1). The number "24″ is the total number of energy line shapes (i.e. four scattering channels for six reflections) fitted simultaneously.